\documentstyle[aps,graphicx,floats,prl,amsmath]{revtex}

\begin{document}
\title{Kinetic Self-Assembly of Metals on Co-polymer Templates} 

\author{ Ajay Gopinathan }

\address{Dept. of Physics and James Franck Institute,
The University of Chicago,  Chicago, IL 60637}
\date{\today}

\maketitle

\begin{abstract}
   In this work we seek to understand some of the fundamental processes that govern self-assembly at the nanoscale in the context of the formation of metallic structures on patterned co-polymer templates. To this end we focus on the experiments  conducted by Lopes {\it et al} \cite{ward1}, where morphologies resulting from the evaporation-deposition of different metals on PS-b-PMMA phase separated templates were studied. We show that the different morphologies obtained can be understood in terms of the relative importance of the energetics and kinetics. We then focus on a particular morphology: micron long wire-like states obtained by the evaporation-deposition of silver on the template. We show the existence of ``non-trivial'' correlations  between adjacent wires that can be understood based on a purely kinetic mechanism. We also compare these correlations quantitatively to those obtained from simulations done with the relevant experimental parameters and find them in good agreement.

\end{abstract}

\section{Introduction}
The field of nanotechnology has seen an explosion of interest over the last decade or so. This has given a big boost to the study of the basic physical phenomena that occur at these small length scales which can be broadly categorized into quantum size effects and interfacial phenomena. The fabrication of  materials that are structured at the nanoscale also poses a formidable challenge if we are restricted to conventional lithographic techniques. This has brought to the forefront approaches that use self-assembly to produce ordered structures spontaneously, mostly relying on interfacial and entropic effects, as in the phase separation of diblock co-polymer systems~\cite{fredrickson}. A natural extension was then to use these self assembled sytems that have structure on the nanoscale as a template, since they have spatially modulated surface properties, to self-assemble the next level of structures \cite{ward1,ward2,zehner1,zehner2,bardotti}.\par
 In this work we examine an experimental realization of the above methodology \cite{ward1}, where the authors used PS-b-PMMA (Polysterene-Polymethylmethylacrylate ) as a template onto which they deposited various metal films. One might expect that the high surface energy of the metals might dominate the interactions, thus rendering the template irrelevant. However the authors observed a variety of structured morphologies, some that respected the template and others that did not. We show that one can understand the different morphologies based on the relative importance of the kinetics and energetics under the given conditions. We also show that the nature of the kinetics can lead to interesting non-trivial correlations developing in the structures.\par
\begin{figure}
  \centering
  \includegraphics[width=5.0in]{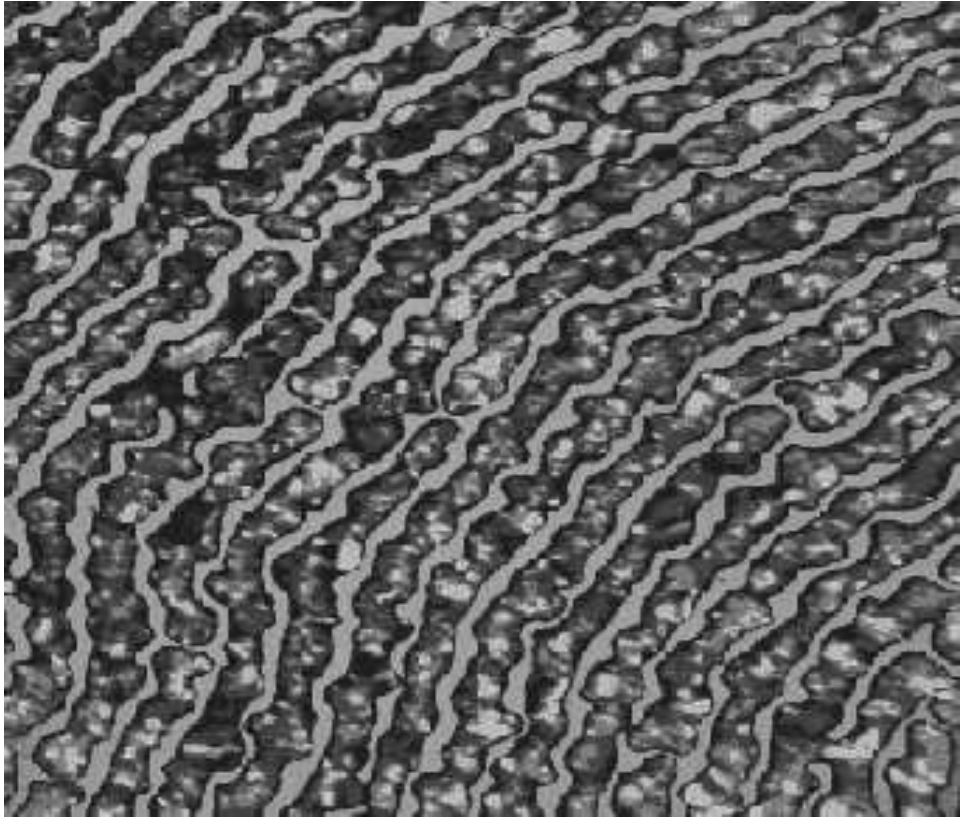}
  \vspace{1ex}
  \caption[Silver nanowires on PS-b-PMMA template]{TEM image of silver nanowires on PS-b-PMMA template. The darker contiguous structures bounded by a dark border are the silver wires and the lighter regions inbetween are PMMA regions not covered by the silver. PS domains are underneath the silver wires. Different shades of gray in the wire regions show different nano-crystalline orientation. Courtesy of W. Lopes and H. Jaeger.}
  \label{fig:nanowire}
\end{figure}

 We now briefly review the experimental procedure followed by Lopes {\it et al} \cite{ward1,ward2}. The template was prepared by first spin coating an ultrathin film of PS-b-PMMA diblock copolymer from solution onto a substrate, which was then subjected to annealing in an inert atmosphere to induce phase separation of the two blocks, producing a fingerprint-like pattern with the two blocks alternating along the film surface. The metals were then evaporated and deposited at room temperature in an inert Argon atmosphere at $10^{-6}$ Torr . Fig. \ref{fig:nanowire} shows typical structures obtained with 120 \AA ~  of silver as deposited. It depicts the formation of micron-long contiguous wire like structures that run along one of the domains (PS). These structures are remarkably stable (given the high surface tension of silver) and do not break up over a timescale of months.\par
\begin{figure}
  \centering
  \includegraphics[width=5.0in]{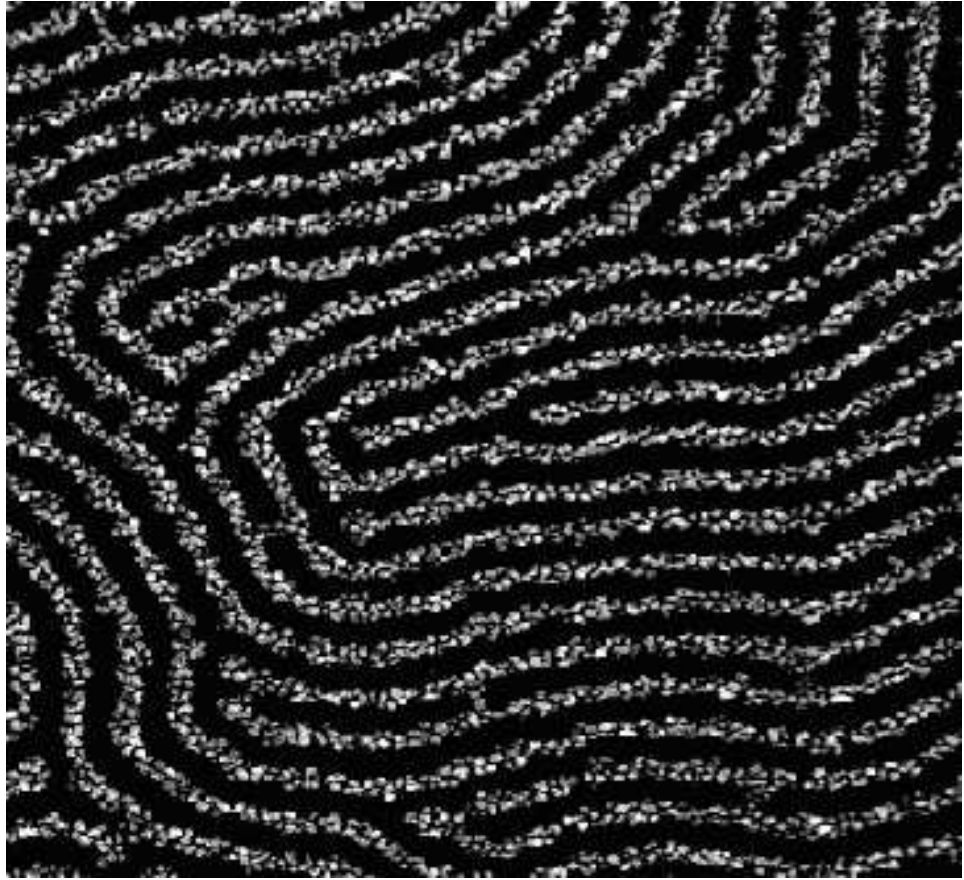}
  \vspace{1ex}
  \caption[Gold nanochains on PS-b-PMMA template]{TEM image of gold nanoclusters on PS-b-PMMA template. The bright spherical structures are the gold nanoclusters  which lie on the PS domain. It is to be noted that the gold is present both at the surface and in the bulk of the PS domain in the form of spherical clusters. Courtesy of W. Lopes and H. Jaeger.}
  \label{fig:nanochain}
\end{figure}

 Following the same procedure for gold, however, does not yield the same results. The gold does not show a high selectivity for the PS domain, but forms clusters in both domains at low coverages and eventually as the coverage is increased, forms large amorphous blobs which ignore the template. To obtain a pattern respecting the template, the authors performed repeated annealing steps (at $180^{o}$ C for 1 min) between depositions of 10 \AA ~  gold to induce the diffusion of the gold to the energetically favorable PS domains. Fig.\ref{fig:nanochain} shows a typical pattern obtained for 60 \AA  ~ gold. Note that in contrast to the case with silver no contiguous wire-like structures are formed. Instead, we have  ``nanochains'' of densely packed gold clusters in one domain. It is our intention in this work to understand the energetics and kinetics that govern the self-assembly process and account for the different morphologies.\par
This paper is organized as follows. In section 2 we show that the wire-like states are non-equilibium states by explicitly computing the free energy. In section 3 we review recent work \cite{combe,jensen2,lewis2} on the kinetics of nanocluster coalescence and estimate the relevant timescales in this particular problem. Based on these estimates, we can account for several experimentally observed features including those in other experiments, and the different morphologies encountered. In section 4  we introduce our methods of image analysis that we use to compute the width distribution of the ``river'' regions between adjacent silver wires as well as the correlations between adjacent silver wires. In section 5 we show how to qualitatively understand these correlations and present our simulations. Finally we compare the simulation results quantitatively, using the river width distribution and correlations, to the experimental results. In section 6 we present our conclusions and scope for future work.

\section{Energetic Considerations}
\begin{figure}
  \centering
  \includegraphics[width=5.0in]{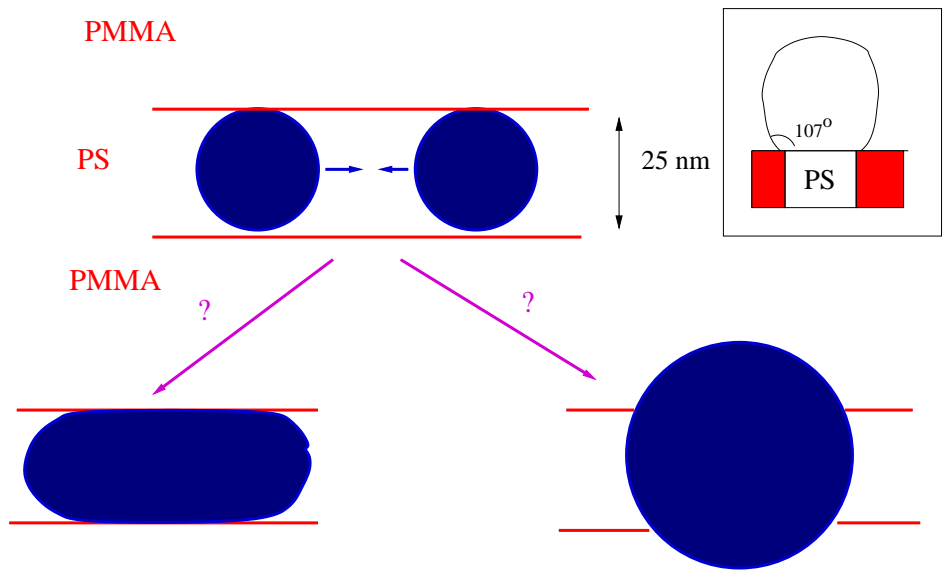}
  \vspace{1ex}
  \caption[Schematic of nanocluster coalescence]{Schematic view of two spherical metal droplets coalescing and their possible final states. The dark regions shown represent the area of contact of the three dimensional droplets with the surface. Inset shows a side-view of a pre-coalescence spherical droplet in contact with the substrate. The PS phase is directly beneath and the two shaded domains on either side are PMMA domains. The contact angle shown corresponds to that obtained from the interfacial energies quoted in the text. }
  \label{fig:draw2}
\end{figure}

  We will now ascertain whether the long wire like states of metal described earlier can be equilibrium states. Consider two blobs of metal (silver/gold) that come into contact, as shown in fig.\ref{fig:draw2}. The blobs are assumed to be spherical caps (spherical balls with  flat bases) with the contact angle being determined by the relevant interfacial energies. The question that we ask is this: do these blobs come together to form a larger blob, or is  the difference in interfacial energies on the two phases sufficient to constrain it to  form a cylindrical blob (see fig.\ref{fig:draw2})? \par

To answer this question we first need the relevant interfacial energies. By far the largest of these is at the metal/vacuum interface. The surface  energy for a clean gold surface is about 1.35 J/m$^{2}$ (1.3 J/m$^{2}$ for silver) \cite{goldPS,augam,aggam}. We will neglect any orientation dependence of the surface energy in these estimates. It is to be noted that impurities,  or reaction with the atmosphere, can drastically affect these energies, reducing it , in some cases, by almost an order of magnitude \cite{redgam}. However, since these experiments were done in an inert atmosphere obtained by drawing a high vacuum and then backfilling with inert Argon up to a pressure of about $10^{-6}$ Torr, we do not anticipate these effects~\cite{wardcom}. In contrast the  surface energy for the polymers is very small (about $3-4 \times 10^{-3}$ J/m$^{2}$ )~\cite{polgam}. The interfacial energy for gold-polystyrene (Au/PS) has an intermediate value of about 0.1 J/m$^{2}$ \cite{ward2}. From the difference in mobilities on PS and PMMA, one can estimate the difference in the Au/PS and Au/PMMA interfacial energies. This is done by using the relation $m ~ \exp{-\Delta E / kT}$~\cite{ward2}, where  $m$ is the ratio of mobilities and $\Delta E$ is the energetic difference between the two phases for a single adatom. The Au/PMMA interfacial energy is calculated to be approximately 0.3 J/m$^{2}$. That there is a difference is not surprising since there is strongly suggestive experimental evidence that shows that the interactions at the Au/PS and Au/PMMA surfaces are qualitatively different \cite{rong}. \par
We will now compute the free energies of the two configurations of fig.\ref{fig:draw2} : the cylindrical blob and the spherical blob. Here we assume that only the metal deforms and the substrate remains undeformed. Given that the surface energy of the metal is dominant, one might intuitively expect the spherical final state to be the equilibrium state. We now show this explicitly. We denote by $\gamma_m$ the surface energy of the metal and by $\gamma_{m,PS}$ ($\gamma_{m,PMMA}$) the interfacial energy of the metal PS (PMMA) interface. Since the surface energies of the free polymer are so small by comparison, we neglect them as a first approximation. We denote by $A^{c}_{m,PS}$ ($A^{c}_{m,PMMA}$) the area of contact between the metal and PS (PMMA) in the cylindrical final state and by $A^{s}_{m,PS}$ ($A^{s}_{m,PMMA}$) the same areas in the spherical final state. Note that $A^{c}_{m,PMMA}=0$ (see fig.\ref{fig:draw2}). We now denote by $A^{c}_{m}$ ($A^{s}_{m}$) the free surface area of the metal in the cylindrical (spherical) final state.
The total energy in either final state is then given by
\begin{equation}
  E_{c/s} =\gamma_m A^{c/s}_{m} + \gamma_{m,PS} A^{c/s}_{m,PS}+ \gamma_{m,PMMA} A^{c/s}_{m,PMMA}
\end{equation}
Taking the spherical blob to be a spherical cap and the cylindrical blob to be a pill shaped object with a flat base and utilising the dimensions in fig \ref{fig:draw2} and the contact angles derived from the values of the interfacial energies we can compute all the relevant areas. Using these values in the above expression yields
\begin{equation}
  E_{c} - E_{s} \approx 10^{4} kT
\end{equation}
with T being room temperature. So we see that the surface tension of the metal indeed dominates and the equilibrium state of two coalescing nanoclusters is the spherical state. Thus the long wire like states appear to be truly non-equilibrium structures. We are now led to ask what the typical timescales are for the equilibration of two coalescing nanoclusters.

\begin{figure}
  \centering
  \includegraphics[width=5.0in]{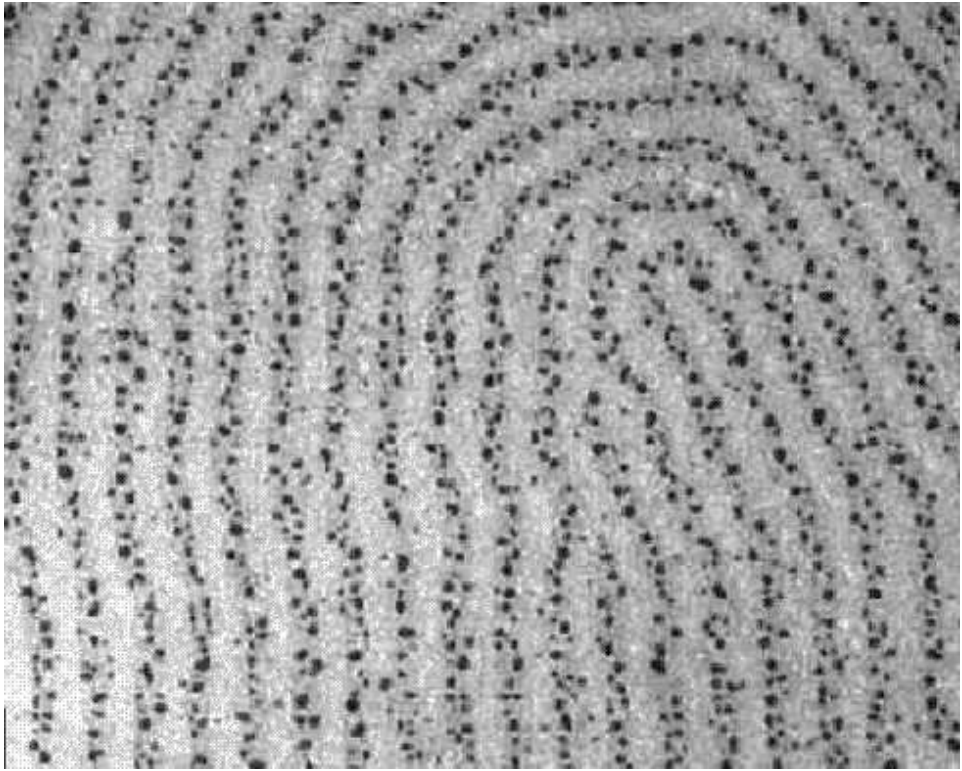}
  \vspace{1ex}
  \caption[Decoration pattern for 25 \AA ~of evaporated silver]{TEM image of decoration pattern for 25 \AA ~of evaporated silver. The silver appears as dark spherical clusters. The domain repeat spacing is about 25 nm. Note the almost all the clusters are in the PS domains. Courtesy of W. Lopes and H. Jaeger.}
  \label{fig:25AAg}
\end{figure}

 \begin{figure}
  \centering
  \includegraphics[width=5.0in]{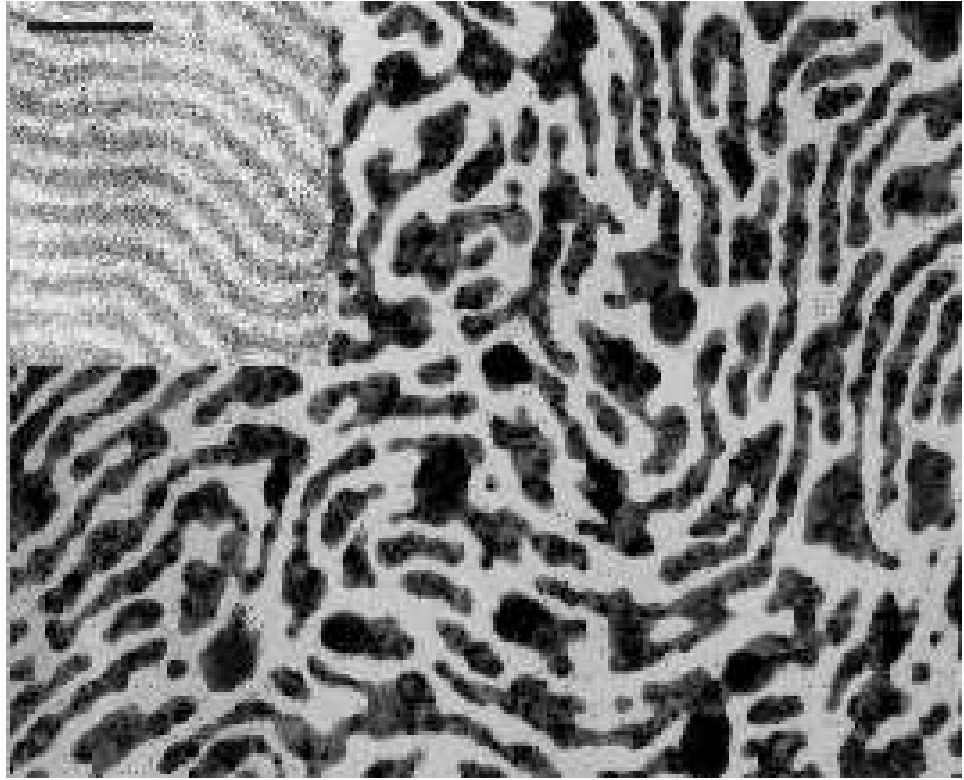}
  \vspace{1ex}
  \caption[TEM image of 120 \AA ~silver sample after annealing]{TEM image of 120 \AA ~silver sample (as in fig.\ref{fig:nanowire}) after annealing at $90^{o}$ Celsius for 3 hours. Scale bar is 200 nm long. Notice that the wires have broken up and are beginning to ``ball up'', forming larger single crystal domains (larger areas of the same shade of gray ) and ignoring the template. The inset shows a TEM image of a similar sample after removal of the silver with gold etch indicating that the underlying template has not been affected by the annealing. Courtesy of W. Lopes and H. Jaeger.}
  \label{fig:annealed}
\end{figure}

\section{Kinetic Considerations}
 The time it takes for metal particles to coalesce, or sinter, is a vital factor in determining the final morphology in growth by evaporation deposition. Conventional approaches to this problem are along the lines of the classic treatment by Herring, Nichols and Mullins \cite{herring,nichols}. The dominant mass transport mechanism is taken to be surface diffusion driven by gradients in chemical potential which depend on the surface curvature. This kind of treatment assumes that a continuous description is feasible which is possible only when the object under consideration is large enough and has a disordered or rough surface as for a liquid droplet. With these assumptions the authors show that the equilibration time $t_{eq}$ scales with the linear dimension $L$ of the particle as $L^{4}$, or in other words, as $N^{4/3}$, where $N$ is the number of atoms. However, for the case of metal nanoparticles at room temperature (below their roughening temperature), it is difficult to see why the above results should still hold.\par
 Recently Combe {\it et al} \cite{combe,jensen2} addressed this very question. They performed Kinetic Monte Carlo simulations of metallic nanoclusters and studied the kinetics of equilibration. They found that at high temperatures close to or above the melting point of the clusters  the continuous approach above produces the correct exponent of $4/3$ (1.33). However, as the temperature is lowered, the relaxation time remains consistent with a power law scaling with $N$ but with  the exponent increasing dramatically to about $5.4$ at $400$ K. They show that the rate-limiting step changes from surface diffusion to the nucleation of new terraces of atoms on facets that appear as the temperature is lowered below the roughening temperature. It is to be noted that the sintering of two contacting grains is not limited by terrace nucleation and should thus be fast.\par
 We will now estimate the relevant time scales in our problem using the phenomenological law
\begin{equation}
 t_{eq} \sim N^{q}
\end{equation}
 with the values for the temperature-dependent exponent $q$ obtained in \cite{combe}. For a metal (gold) cluster with about $N_0 = 10^3$ atoms at $800$K, the equilibration time is about $100$ ns \cite{lewis2}. In comparison, the prediction from the macroscopic theory gives a time an order of magnitude lower even for a relatively small cluster such as this. We take this timescale, $\tau_0 = 100$ ns to compute the the equilibration times for larger clusters. As a first approximation we neglect the effect of temperature on $\tau_0$, i.e., we assume that a $1000$ atom cluster takes roughly the same time to equilibrate even as the temperature is lowered, which will be increasingly true as the cluster size is reduced.\par
 We next look at a typical cluster in our problem. Fig \ref{fig:25AAg} shows typical clusters that are obtained at low coverages. Given that the width of the PS domain is about 25 nm, we see clusters that are about 10-12 nm in diameter. Two of these that are beginning to coalesce  will contain about $N = 10^5$ atoms. The equilibration time, at $T=400$K for example, can now be computed using the scaling law.
\begin{equation}
 t_{eq} = \tau_0  (N/N_0)^{q}
 \end{equation}
Using the values $N=10^5$, $N_0 = 10^3$ and $q=5.4$ at $T=400$K we get $t_{eq} \approx 2$hrs. Thus, even at $400$K, the equilibration time can run into hours. Extrapolating the value of $q$ \cite{combe} to room temperature can give values as high as $q \approx 7$, which translates into equilibration times of many {\it months}. The two other timescales that are relevant are the time it takes for two typical clusters to find each other ($t_f$) and the timescale of the deposition process. The deposition process typically takes a few minutes and is hence a much smaller timescale. Using typical values for cluster diffusion constants in such situations \cite{lewis1,shull1,jensen1} , $D \approx 10^{-14}$ cm$^2$/s and a typical separation, $l$, of the order of nanometers, we may estimate $t_f$
\begin{equation}
 t_f \sim l^2/D \approx 1 sec
\end{equation}
Hence $t_f \ll t_{eq}$. \par

Since the coalescence time is much greater than the typical time for the clusters to find each other, we must necessarily anticipate the growth of extended non-compact structures. In the case of the silver being deposited on the PS-b-PMMA template, the difference in mobilities of the silver on the PS (low) and PMMA (high) implies that the silver is quickly made available over the energetically favorable PS domain where it tends to stay. This in turn implies that the non-compact structures we anticipate will lie along the PS domains, resulting in, as observed experimentally, long wire-like states on the PS phase (as in fig \ref{fig:nanowire}). The argument above is quite general and should apply to any metal that shows sufficient overall moblity and energetic preference for one of the two phases. However, Lopes {\it et al} observed that for gold, following the same procedure did not form wires. They found that gold as evaporated and deposited at room temperature did not show sufficient selectivity which led to the occurence of large clusters on the PMMA phase which were almost immobile. Further deposition did indeed lead to non-compact objects but which had no directional preference, thus forming patterns that ignored the template. Lopes {\it et al} tried to obviate this problem by annealing at 250 Celsius in between deposition steps so as to get the gold to diffuse to the energetically favorable PS domain. From our arguments about the kinetics we presented, it becomes clear that doing so will dramatically decrease $t_{eq}$, resulting in equilibrium spherical structures. Indeed, this is what they observed, as shown in fig. \ref{fig:nanochain}. The situation was further complicated by the fact that during annealing gold also diffuses into the bulk, reducing the amount of gold at the surface.\par
 Another interesting observation is that the micron long wire-like states of silver, when annealed at a slightly elevated temperature (90 Celsius) for about 3 hours, break up  and ``ball up'', ignoring the template as in fig.\ref{fig:annealed}. This temperature is below the glass transition temperature of the substrate and the melting temperature of the silver. Thus, this phenomenon seems to point to the fact that these structures are indeed non-equilibrium states. From our estimates of the relevant timescales the reason for this phenomenon becomes apparent. Raising the temperature to about 400 K changes the equilibration time $t_{eq}$ from months to hours. Therefore waiting at the elevated temperature for a few hours will indeed result in observable changes in the morphology as the system moves toward equilibrium.\par

 Now that we have a qualitative understanding of how these wire-like states are formed, we will focus on certain characteristics of these states. We will show that there exist non-trivial correlations between adjacent wires and further, that these can be understood as arising from a kinetic mechanism.
\section{Correlations}
 Looking closely at the silver wires (fig.\ref{fig:nanowire}) one suspects the presence of correlations between adjacent silver wires. If we designate by the term ``river'', the silver free PMMA regions in the picture and by the term ``bank'', the boundary of the silver-PMMA interface, then an outward bulge on one bank seems to correspond with an inward  depression on the opposite bank. It is to be noted that the underlying PS-PMMA interface is relatively smooth, so that the bumps are merely on the silver structures. The presence of correlations, if true, is significant because it might imply that there is an interaction between the banks whose origins may be electrostatic or even mechanical (strain mediated by the substrate). This might be an alternate explanation for the stability of the wires. We will show, however, in the next secton that the observed correlations can be accounted for simply by a kinetic mechanism. We now attempt to measure these correlations to see if they indeed exist and to quantify them if they do.\par
As a first attempt to quantify the correlation we look at the simplest statistic which could yield some information, namely, the river width distribution.
\begin{figure}[!p]
  \centering
  \includegraphics[width=5.0in]{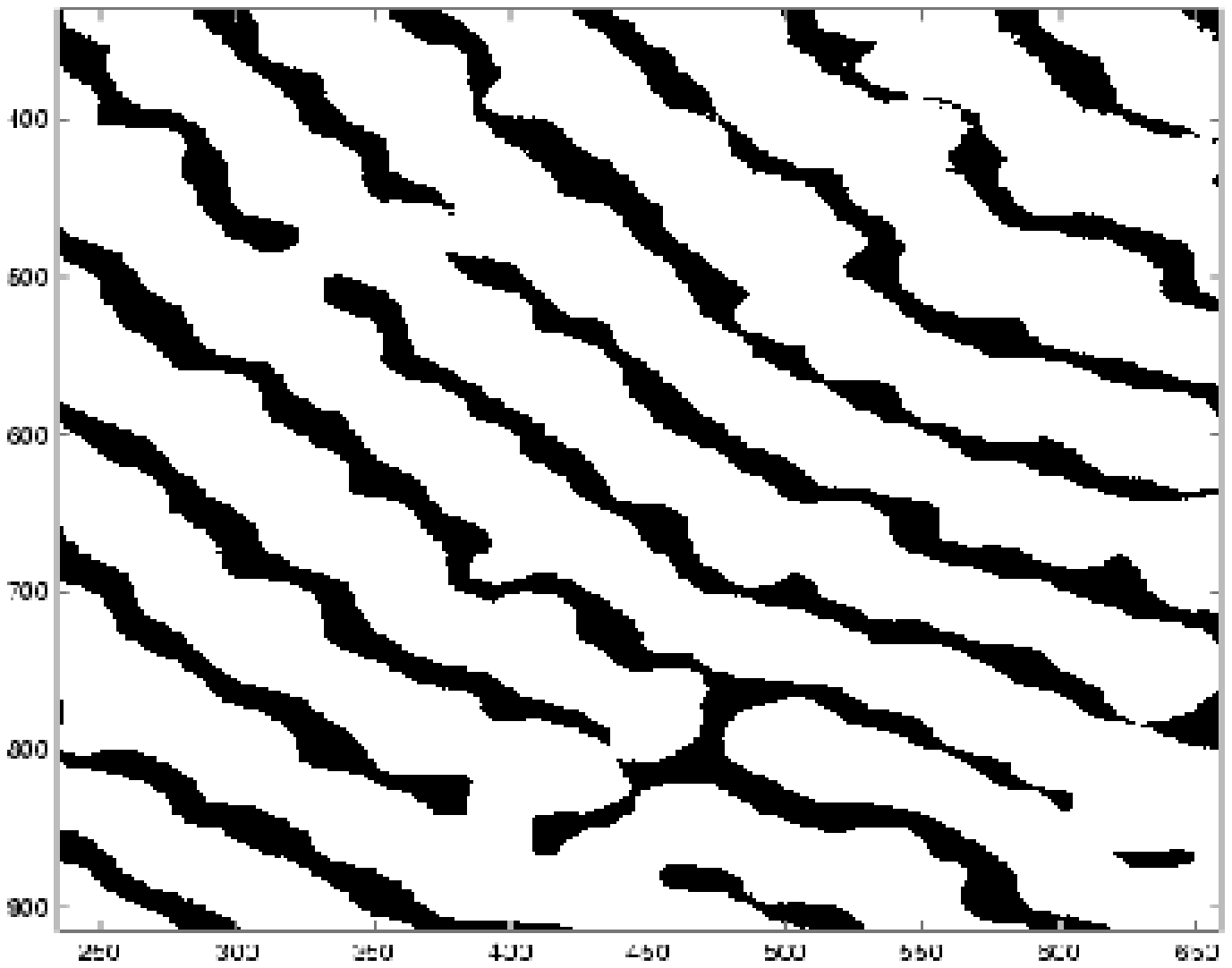}
  \vspace{1ex}
  \caption[Processed binary image of nanowires]{Representative binary image obtained by processing a TEM image like fig.\ref{fig:nanowire}. The lighter regions have pixel value 1 and represent regions where silver is present (the ``wires''). The darker regions have pixel value 0 and represent the exposed PMMA ``river'' regions.  }
  \label{fig:bwwire}
\end{figure}

\begin{figure}
  \centering
  \includegraphics[width=5.0in]{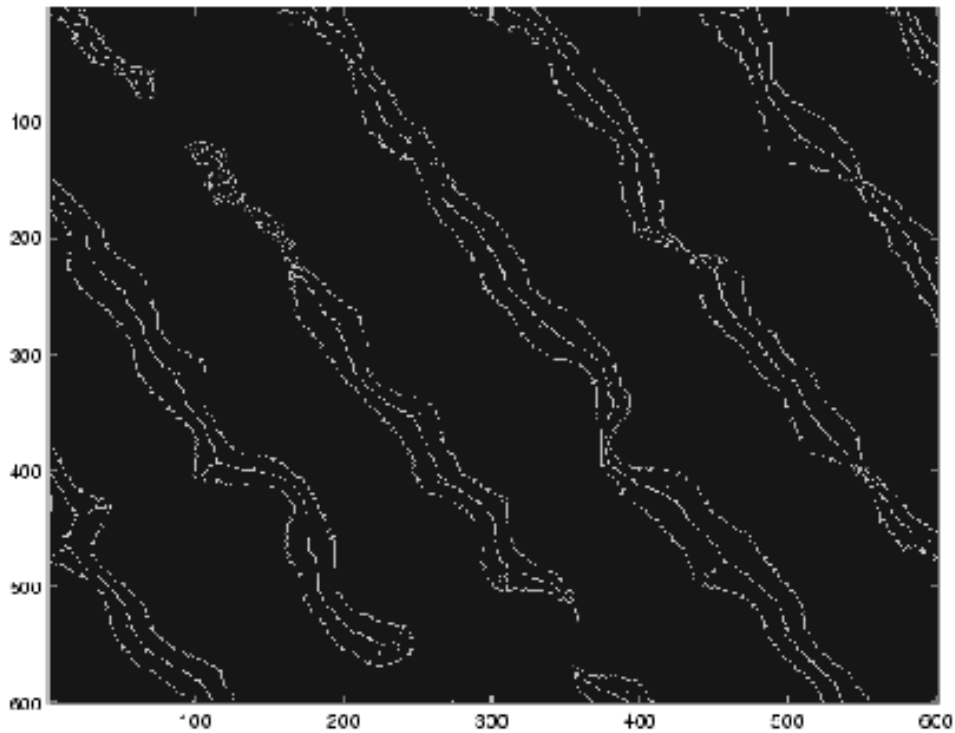}
  \vspace{1ex}
  \caption[Midpoints of the ``river'' regions]{Midpoints of the rivers. White lines denote the banks of the rivers and the lines of midpoints running down their centers. }
  \label{fig:postbao}
\end{figure}

\begin{figure}
  \centering
  \includegraphics[width=5.0in]{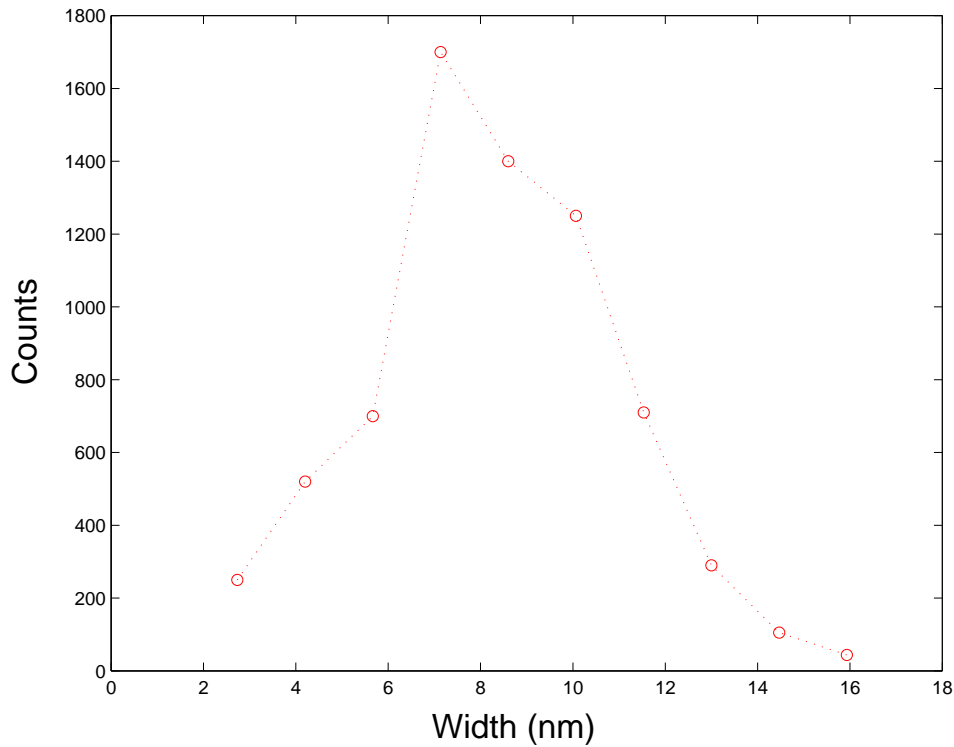}
  \vspace{1ex}
  \caption{ Width distribution obtained from the experimental TEM images.}
  \label{fig:exptwd}
\end{figure}

We first map the TEM images (see fig. \ref{fig:nanowire}) to an image with binary values for the pixels. A pixel has a value 1 if the corresponding point in the image has silver and a value 0 if it does not. We are then left with an image that is shown in fig. \ref{fig:bwwire}.  The rivers are now pixels with value zero. We now wish to ascertain the midpoints of the river, i.e., points that are ``equidistant'' from either shore. It is not {\it a priori} obvious how one can rigorously define the midpoints of the river regions since the rivers curve (following the underlying PS-PMMA domains) thus making it difficult, given one point on one bank, to pick out a corresponding point ``across'' the river. We tackle this problem as follows. For each point $P$, in the binary image with pixel value zero (a point in the river), we find the closest point on either bank. We denote the distance by $r$. Next we compute the average pixel value in a circle of radius $2r$ around point $P$ and assign this value to the point. Now it is to be noted that the value assigned increases as the point moves further from one bank. It reaches a maximum somewhere in the middle and then begins to decrease as the oppposite bank is approached. This is done for all points with pixel value zero. We thus generate an image with all the pixels in the river regions having non-zero values assigned to them, with values being largest in the middle of the rivers. By appropriate thresholding we can get a line of points running down the middle of the river which we denote as the ``set of midpoints'' of the river (see fig.\ref{fig:postbao}). Knowing the value of $r$ associated with each of these points we can plot a distribution of river widths. Fig.\ref{fig:exptwd} shows a distribution obtained from data from four images such as the one in fig.\ref{fig:bwwire}. The average width is 8.24 nm and the standard deviation is 2.6 nm.\par
\begin{figure}
  \centering
  \includegraphics[height=7.0in]{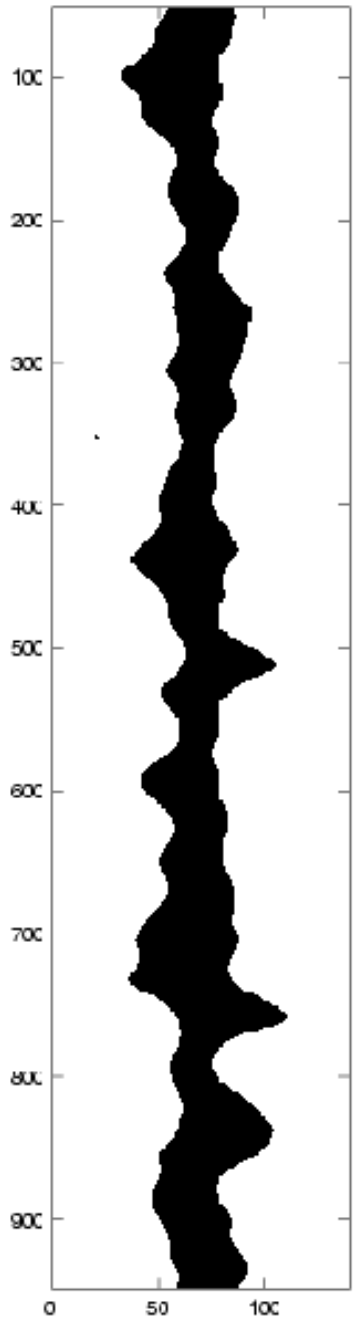}
  \vspace{1ex}
  \caption[Binary image of the control case]{Binary image of the control case. The lighter areas represent regions where silver is present (the ``wires''). The darker area represents the exposed PMMA ``river'' region. }
  \label{fig:contpic}
\end{figure}
\begin{figure}
  \centering
  \includegraphics[width=5.0in]{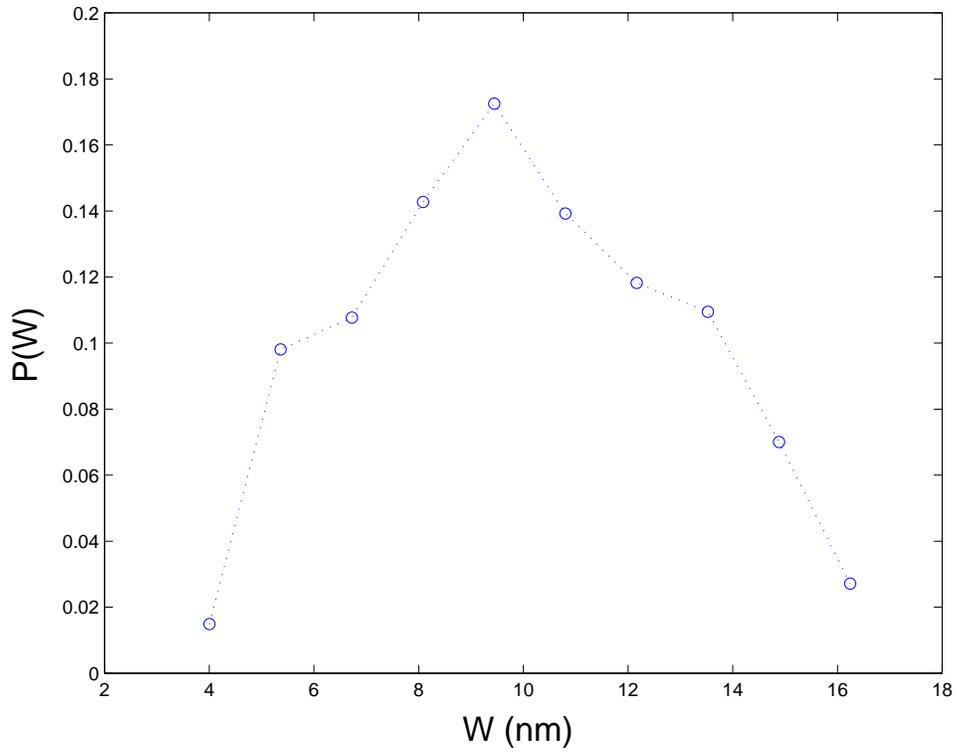}
  \vspace{1ex}
  \caption[Width distribution for the control case]{Width distribution obtained from control case images (like fig.\ref{fig:contpic}).  Probability of a particular width, P(W), is plotted against the width, W, in nanometers. }
  \label{fig:contwd}
\end{figure}
\begin{figure}
  \centering
  \includegraphics[width=5.0in]{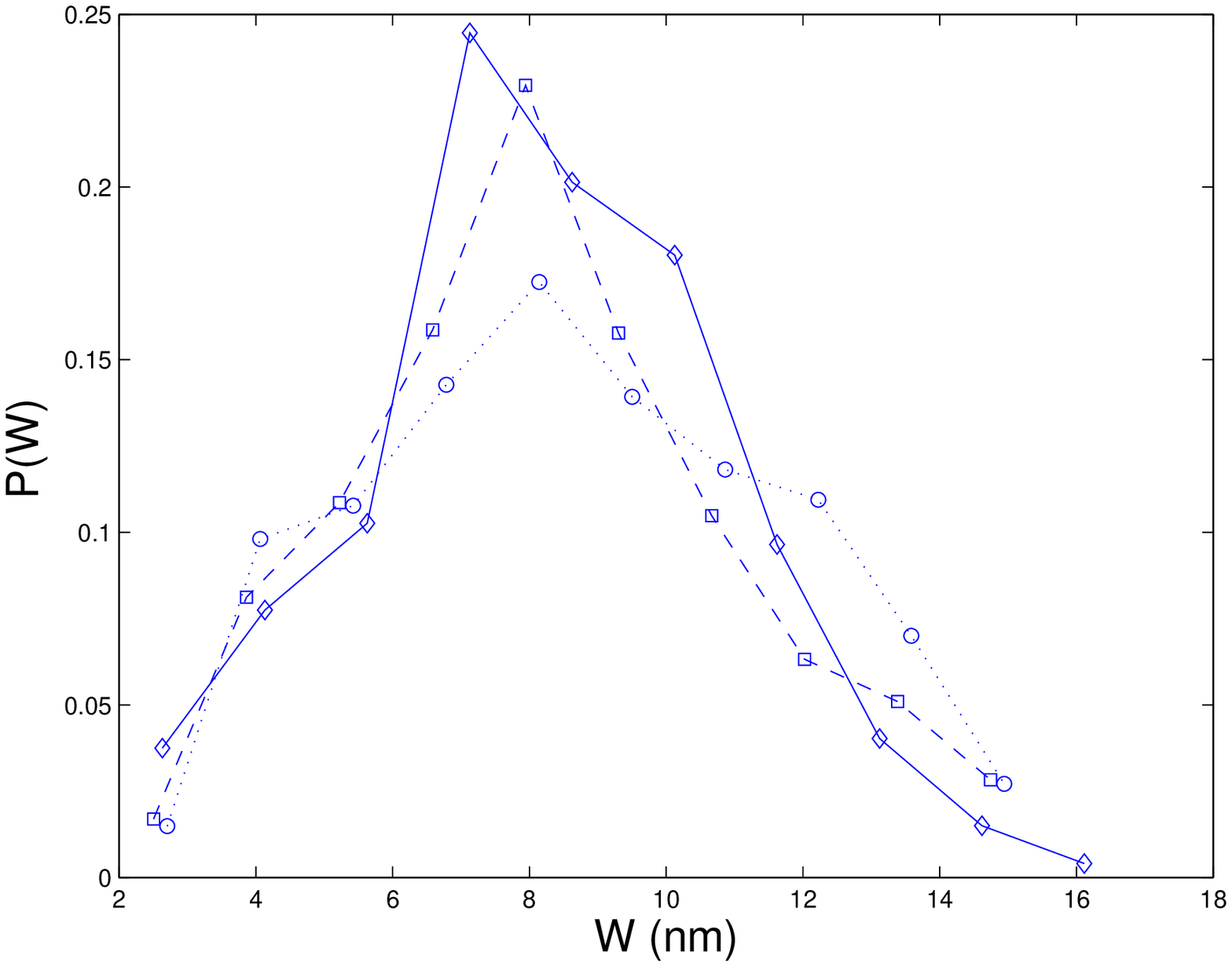}
  \vspace{1ex}
  \caption[Comparison of experimental, simulation and control case width distributions]{ Comparison of the width distributions from three different cases. Diamonds connected by a solid line represent the experimental data. Squares  connected by a dashed  line  and circles connected by a dotted line represent the simulation and control case distributions respectively. For ease of comparison, the distributions have been shifted so that the mean widths coincide. }
  \label{fig:comparewd}
\end{figure}

 We now construct a ``control'' case where the two banks are explicitly uncorrelated. We do this by considering three adjacent strips (see fig.\ref{fig:contpic}) each representing a width of $25$ nm and a length of $500$ nm. The two outer regions represent PS domains and the one in between is a PMMA domain. We then put down spherical silver clusters of a fixed radius at randomly chosen nucleation points  in the PS domains. The clusters are allowed to overlap and by adjusting the density of nucleation points as well as the radius of the spheres we can get contiguous wire-like structures on the PS domains. In reality, the surface tension of silver will tend to smoothen the surface. We model this by a smoothening algorithm which allows material from regions of high curvature to flow to those with low curvature. This algorithm is run till the surface is smooth, i.e., there are no obvious discontinuities in the derivative of the curve representing the silver-PMMA boundary yielding wire-like structures separated by silver free ``river'' regions,  as in fig.\ref{fig:contpic}. The density and radius are chosen so that the ``bumps'' of the wires have roughly the same magnitude as in the experimental situation. It is to be noted that the ``bumps'' on either side (bank) of the silver free PMMA region are uncorrelated by construction. The width distribution of the control case is shown in fig.\ref{fig:contwd}. The average width is 9.53 nm and the standard deviation is 2.7 nm. Though the numbers are not very different, comparing the two width distributions (see fig.\ref{fig:comparewd}) shows that the distribution for the control case is broader and more symmetric while the experimental distribution is narrower with a steep rise and a more gradual fall-off for increasing width. If the two banks meandered in unison one would indeed expect a qualitatively narrower width distribution. Thus the ``narrowness'' of the distribution qualitatively indicates the presence of correlations.\par


\begin{figure}
  \centering
  \includegraphics[width=5.0in]{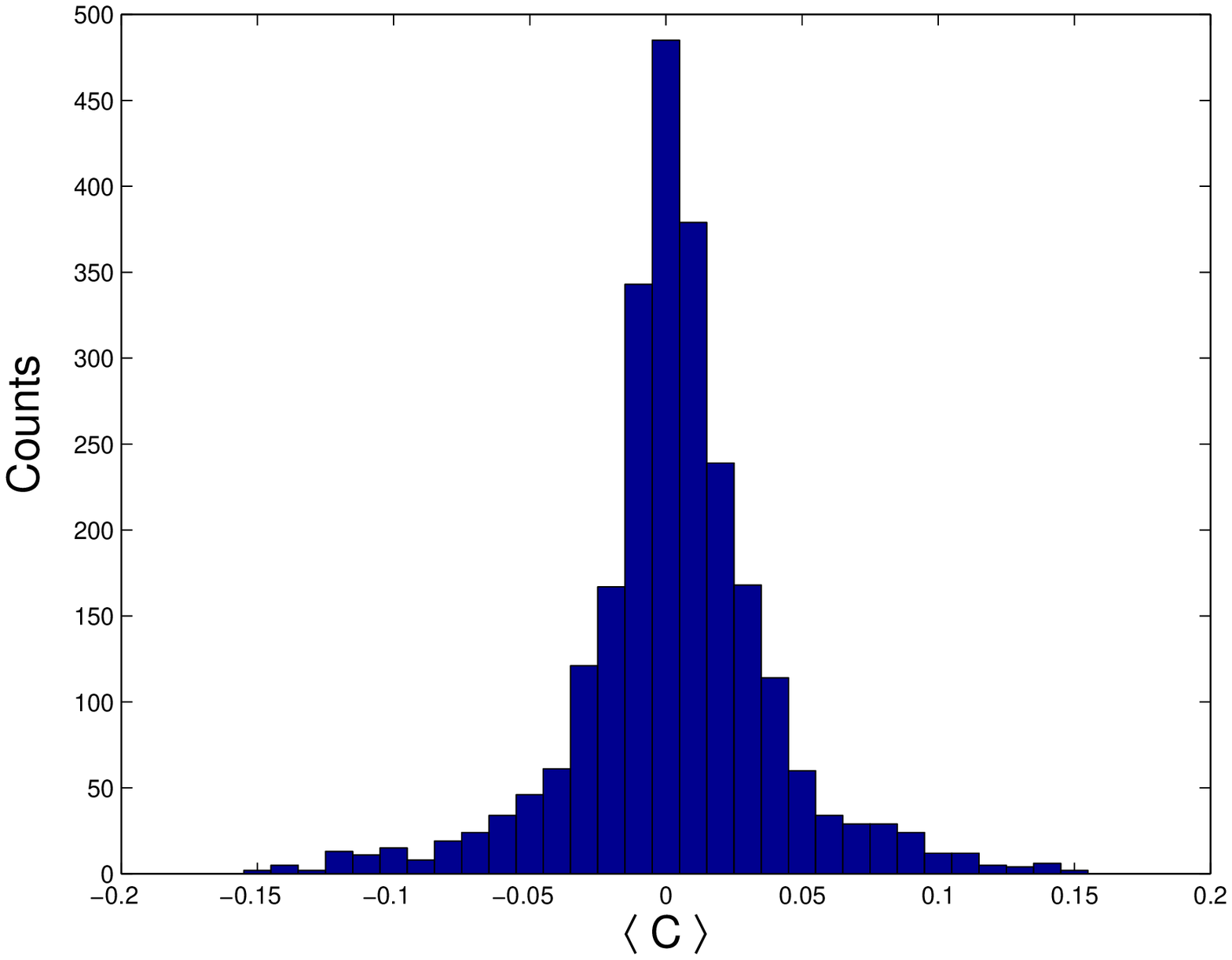}
  \vspace{1ex}
  \caption[Histogram of the correlation statistic $\langle C \rangle$]{ Histogram of the correlation statistic $\langle C \rangle$ obtained from analysing experimental images. }
  \label{fig:chist}
\end{figure}

\begin{figure}
  \centering
  \includegraphics[width=5.0in]{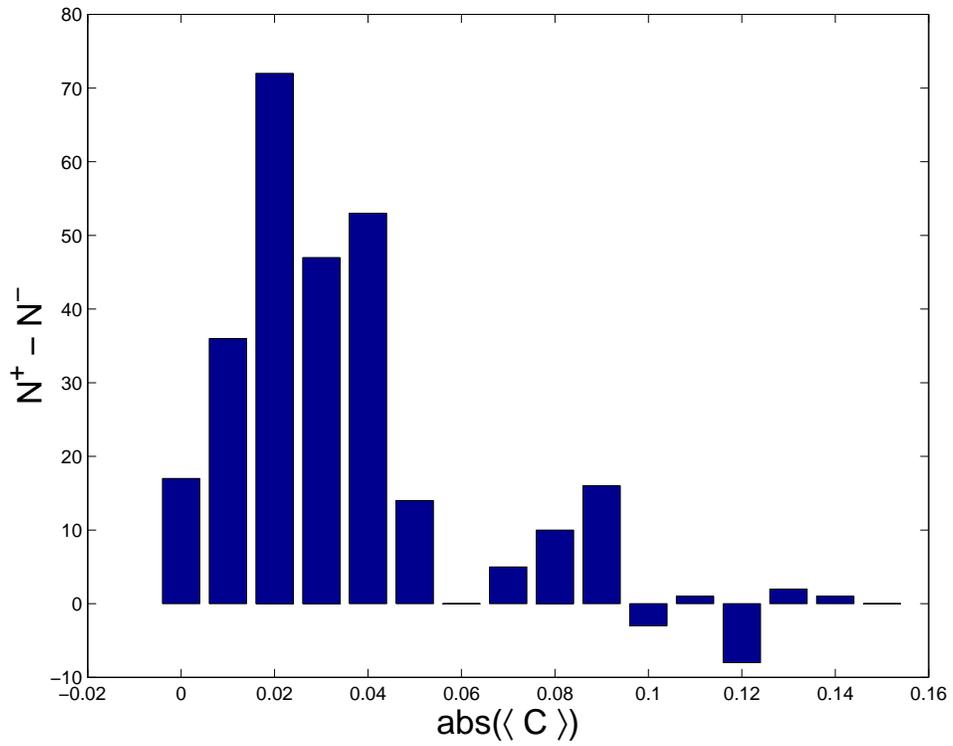}
  \vspace{1ex}
  \caption[Bar plot of the excess of positive values of $\langle C \rangle$]{ Bar plot of the excess of positive values of $\langle C \rangle$. $N^{+}$ is the number of counts in a bin corresponding to a particular value of $\langle C \rangle = +C $ and $N^{-}$ is the number of counts in the bin corresponding to $\langle C \rangle = -C$ in the histogram in fig.\ref{fig:chist}.}
  \label{fig:chistpm}
\end{figure}

 In order to pin down the extent of correlations quantitatively, we now define a new statistic, as follows. For every point in the set of midpoints we identify two corresponding points, one on each bank. This is done by constructing a normal to the local tangent of the line of midpoints. The intersection of the normal with each bank gives us one point, thus yielding a pair of points on opposite banks for each midpoint. We then compute the local curvature  of the bank at each point of this pair and assign to the midpoint, the product of the curvatures. It is to be noted that the product, which we will denote by $C$, will be (a) zero : if the bank on either side is locally straight, (b) negative : if the banks curve away from each other and (c) positive : if the banks curve ``together'' in the same direction. The last case is what we mean when we say ``correlated''. Thus we generate for each midpoint a value based on the correlation of the local curvatures of two banks nearby. We then assign to each midpoint the average value of $C$ of all the midpoints within a distance equal to the local width. Fig. \ref{fig:chist} shows a histogram of the values of $\langle C \rangle$ thus obtained. At first it might appear that positive and negative correlation contributions are the same. However, replotting the histogram with the difference between the positive and negative contributions  at a value of the absolute magnitude of $\langle C \rangle$ versus
$\Vert \langle C \rangle \Vert$  shows a clear trend for the excess of positive contributions (see fig.\ref{fig:chistpm}). 

\begin{figure}
  \centering
  \includegraphics[width=5.0in]{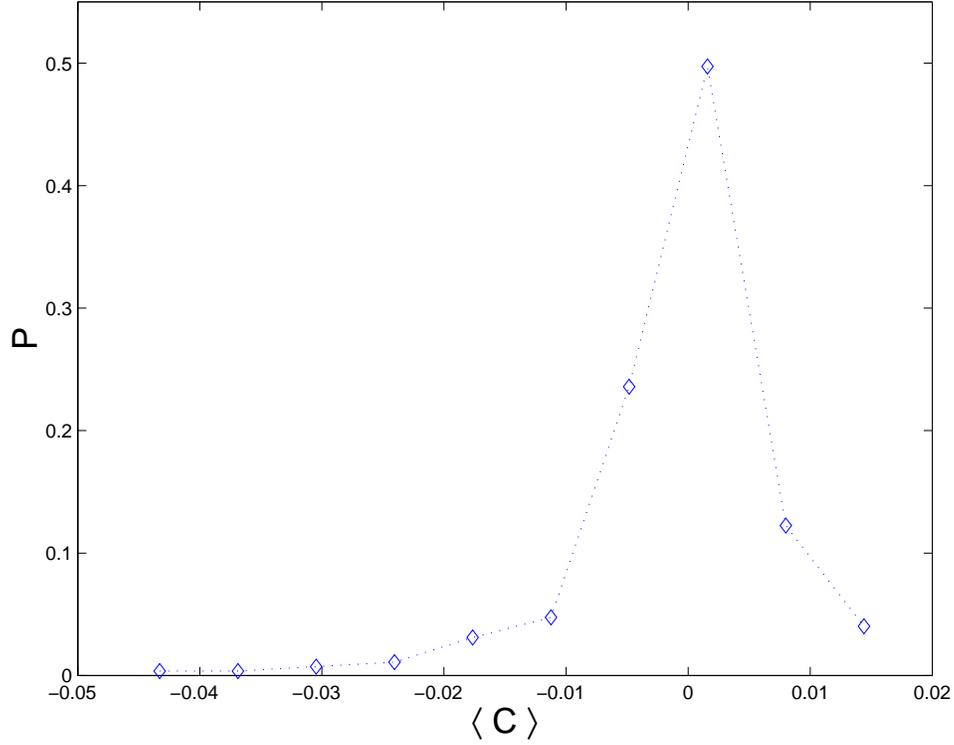}
  \vspace{1ex}
  \caption[Correlation statistic $\langle C \rangle$ for the control case]{ Plot of the distribution of the correlation statistic $\langle C \rangle$ obtained from analysing control case images. Probability of a certain value of the correlation statistic $\langle C \rangle $, P, is plotted against $\langle C \rangle $.}
  \label{fig:conhist}
\end{figure}
\begin{figure}
  \centering
  \includegraphics[width=5.0in]{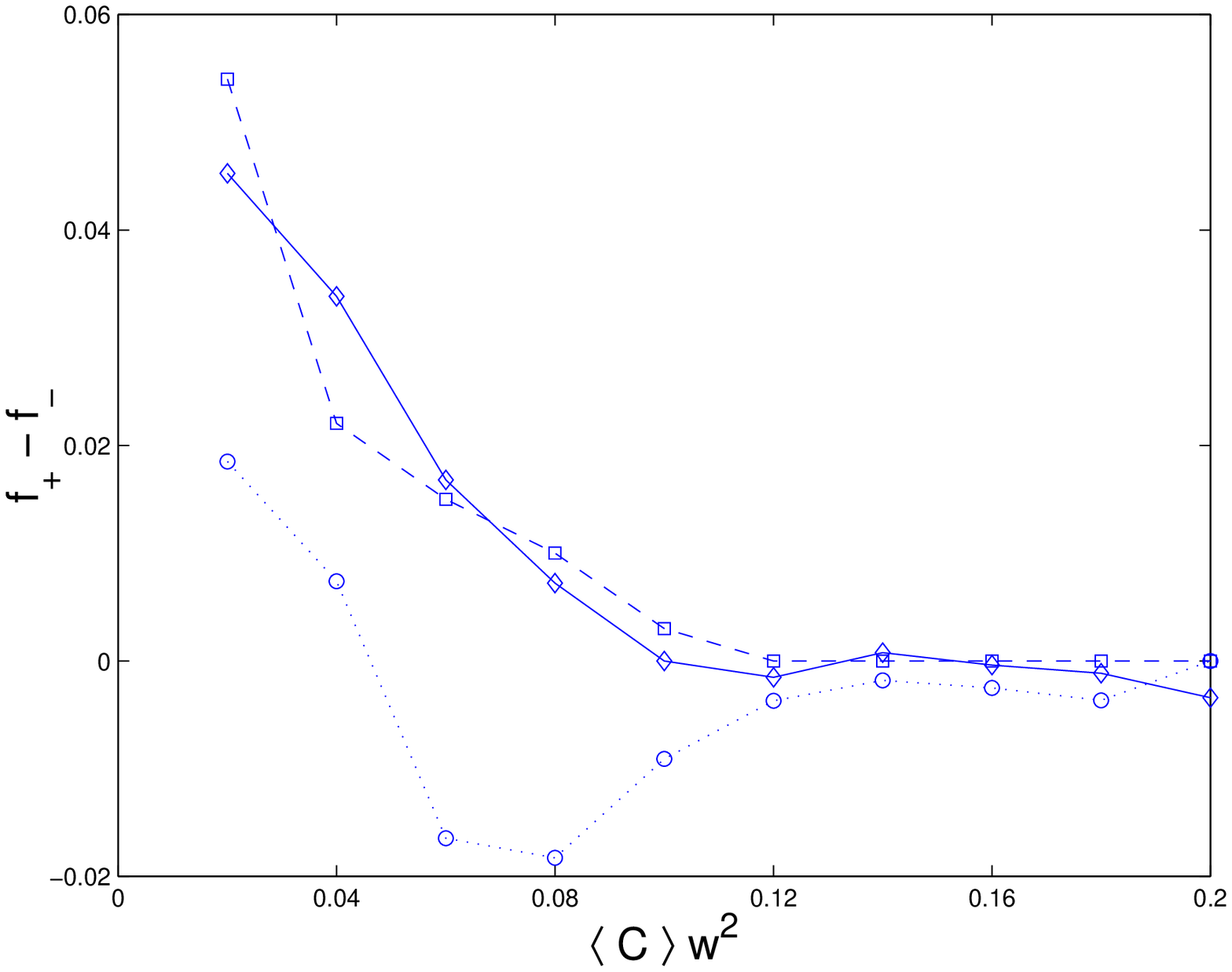}
  \vspace{1ex}
  \caption[Comparison of the excess fraction of positive values of the correlation statistic for the experimental, simulation and control cases]{Comparison of the excess fraction of positive values of $\langle C \rangle w^2$ for the experimental (diamonds with solid line), simulation (squares with dashed line) and control (circles with dotted line) cases. $f_{+}$, $f_{-}$ and $\langle C \rangle w^2$ are as described in the text.}
  \label{fig:compareC}
\end{figure}

Figure \ref{fig:conhist} shows a plot of the same statistic ($\langle C \rangle$) for the control case. In this case there seem to be more negative than positive contributions. Figure \ref{fig:compareC} allows us to compare the difference between the positive and negative contributions for the experimental and the control case. Here $\langle C \rangle$ for each case has been scaled by the square of the average width of the width distribution for that  case, so as to obtain a dimensionless quantity, which can then be compared. The quantity $f_{+}-f_{-}$ is the fraction of data points with a value $+\langle C \rangle w^{2} $ less the fraction of points with a value $-\langle C \rangle w^{2} $. We immediately see a striking qualitative difference between the experimental case with mostly positive values and the control case with predominantly negative values over the same range of values of $\langle C \rangle w^{2}$ indicating that there are indeed non-trivial correlations between the banks.\par
Does the presence of non-trivial correlations between the wires imply a stabilizing interaction between the wires or do these arise simply from the kinetics of the wire formation process? It is precisely this that we will address in the next section.

\section{Accounting for Correlations}
 An important clue is the fact that even for silver, which shows an almost 100\% selectivity at low coverages, there are still some clusters in the PMMA domain (see fig. \ref{fig:25AAg}). This implies that clusters do nucleate and grow in the PMMA domain as well. We later show that simulations with the right diffusion constants do indeed lead to cluster formation in the river regions. Now let us consider a later stage in the process when the basic wire-like states have already been formed, while deposition still continues. Clusters in the river regions will grow due to the incident flux of silver atoms and new clusters will also be formed. As the clusters grow bigger they slow down and explore a more restricted region. Eventually a big cluster will diffuse to one of the two banks in its vicinity thus leading to enhancement of material (a bump) on one bank at the expense of the nearest regions of the opposite bank. Thus this kinetic process leads to the development of correlations between the two banks. We now present simulations of a model of the process to see if we can quantitatively account for the correlations measured in the previous section. \par
We first need to know the diffusion constants for silver clusters in the PMMA region. To do this we refer to the experiments done on the evaporation deposition of silver on homopolymer PMMA and PS \cite{ward2}. For low coverages of silver the author measured both the nearest neighbor distance distribution, as well as the size distribution of the resulting islands at a given flux rate. We performed simulations by considering a square region with periodic boundary conditions that represented a $ 250 \times 250$ nm$^{2}$ section of the homopolymer substrate. We modeled the flux of silver by the introduction of small clusters (about 100 atoms)  at the appropriate rate which were then allowed to diffuse with a predetermined diffusion constant $D_0$. Clusters were allowed to coalesce upon contact, forming larger spherical islands. The diffusion constants of larger clusters were assumed to follow a $N^{-2/3}$ law \cite{ward2,lewis1,jensen1}. By suitably adjusting the value of $D_{0}$, we were able to match both the distributions measured experimentally. This gave us a value of $D_0 = 10^{-13}$ cm$^2$/s for the diffusion constant of the small silver clusters on PMMA. This value is consistent with values for cluster mobility on organic substrates \cite{lewis1,shull1,jensen1}, which are typically of this order. Now that we have fixed the value of the diffusion constant we present our simulations that attempt to capture the correlations. \par
\begin{figure}
  \centering
  \includegraphics[height=7.0in]{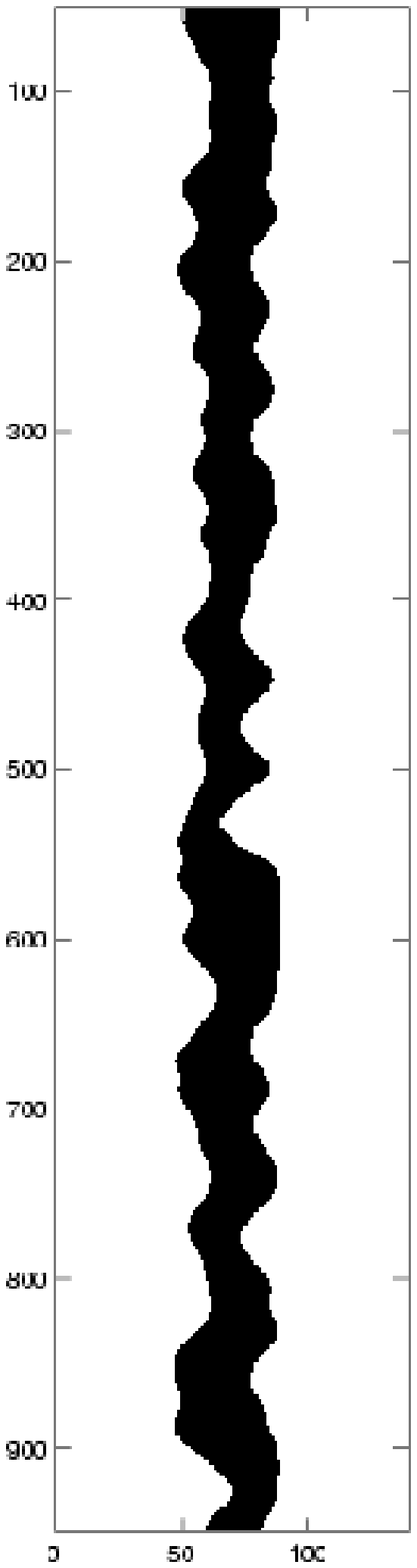}
  \vspace{1ex}
  \caption[Binary image of the ``wire'' and ``river'' regions obtained by simulation]{Binary image of the ``wire'' and ``river'' regions obtained by simulation. The lighter areas represent regions where silver is present (the ``wires''). The darker area represents the exposed PMMA ``river'' region. }
  \label{fig:psmooth}
\end{figure}
\begin{figure}
  \centering
  \includegraphics[width=5.0in]{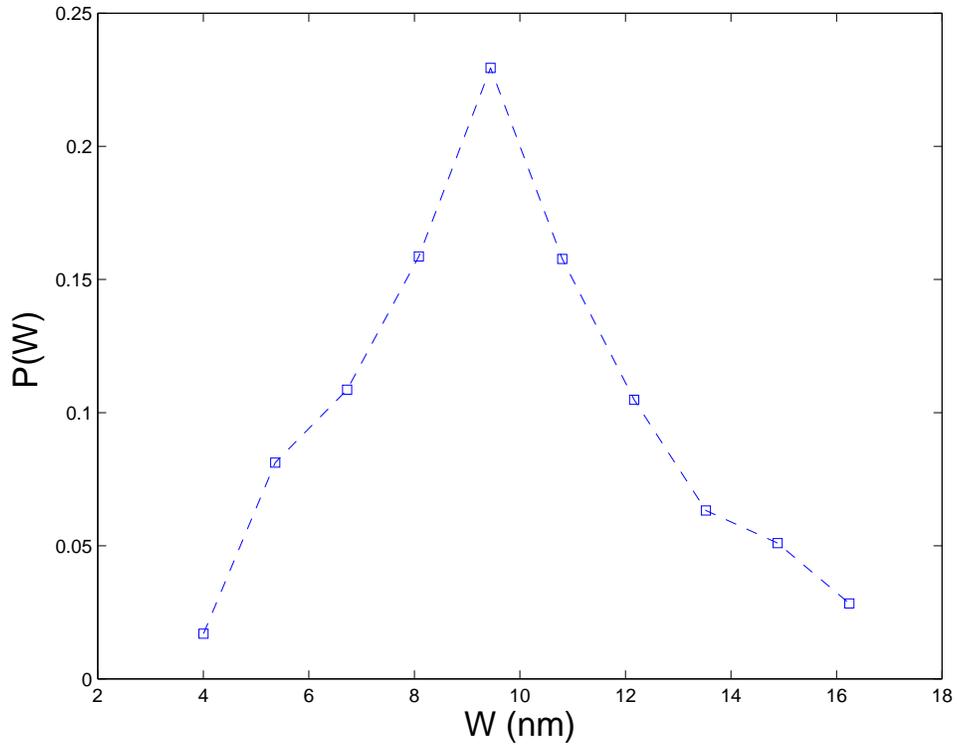}
  \vspace{1ex}
  \caption[Width distribution for the simulation case]{Width distribution obtained from simulation images (like fig.\ref{fig:psmooth}).  Probability of a particular width, P(W), is plotted against the width, W, in nanometers. }
  \label{fig:wdsim}
\end{figure}
\begin{figure}
  \centering
  \includegraphics[width=5.0in]{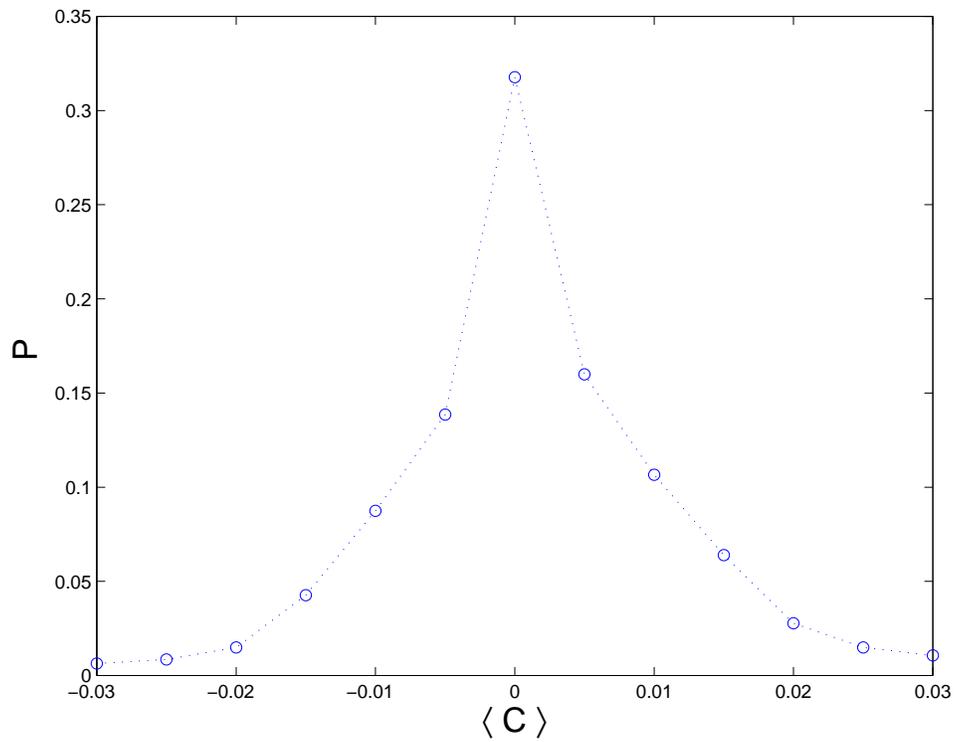}
  \vspace{1ex}
  \caption[Correlation statistic $\langle C \rangle$ for the simulation case]{ Plot of the distribution of the correlation statistic $\langle C \rangle$ obtained from analysing images obtained by simulation (e.g. fig.\ref{fig:psmooth}). Probability of a certain value of the correlation statistic $\langle C \rangle $, P, is plotted against $\langle C \rangle $. }
  \label{fig:chistsim1}
\end{figure}

To this end we consider a rectangular region which represents a $500 \times 25$ nm$^{2}$ section of the substrate. This section represents a PMMA domain of width $25$ nm and length $500$ nm. The boundary conditions are periodic in the long direction. In the short direction, the boundaries are meant to represent a solid silver wall, mimicking the edges of silver wires on the PS domains. Thus a cluster touching these boundaries sticks there and is assumed to be immobile, though it can still grow from the incident flux of silver. The procedure is exactly the same as for the homopolymer case described above. Clusters are incident on the river region at the appropriate flux (set by the experimental value) which then diffuse and coalesce, and eventually stick to the banks on either side. As mentioned before, we do indeed see the formation of large clusters ($\sim 10^4$ atoms) that have not yet been trapped by the walls. We do not know how long the late stage of the process is, so we take a conservative underestimate by continuing the deposition till about 50 \% of the PMMA region is covered. This is to be contrasted with the coverage in the experimental case which is about 66\%. By underestimating the coverage we hope to produce a conservative estimate for the correlations which is still comparable to the correlations in the experimental case. Once the deposition and diffusion process is over we get a series of bumps on both banks of the river region where the clusters have stuck. We then smoothen the surface by the algorithm described in the previous section for the control case. A typical image thus obtained is shown in fig.\ref{fig:psmooth}. A casual inspection shows that the correlations we noticed in the experimental images seems indeed to be present in these images. To quantify this we analyze the simulated images exactly as we did the experimental images and the control case by the method described in the last section. The width distribution obtained is shown in fig.\ref{fig:wdsim}. We notice that it looks very similar to the experimental case (see fig.\ref{fig:comparewd} for comparison) except that it is slightly shifted to the right (larger mean width) due to the underestimation of the coverage. Fig.\ref{fig:chistsim1}  shows the plot for the correlations. The plot looks qualitatively similar to the histogram for the experimental case (fig.\ref{fig:chist}). Fig .\ref{fig:compareC} shows the plots of $f_{+}-f_{-}$ for the experimental, simulation and control cases. We see that the simulation results are in good agreement with the experimental results, both of which are qualitatively different from the plot for the control case. Thus, one can reproduce both qualitatively and quantitatively the correlations observed in the experiment with a simple kinetic mechanism.

\section{Discussion of Results}

We have presented an analysis of the energetics and kinetics that govern the self-assembly process of metals on patterned polymer templates. We have shown that depending on the conditions one can have one or the other dominating the process. Thus tuning the conditions can lead to vastly different morphologies as evidenced by the experiments. In our analysis the only factor that depended critically on the nature of the metals was the selectivity between the two domains. One could then, in principle, form wire-like states of different metals provided there is a sufficient contrast in mobilities and an energetic preference for one of the domains. It is to be noted that achieving this by raising the temperature will have the effect of driving the system to equilibrium (spherical clusters) as seen in the case with gold. This could be done on the other hand by changing the composition of the underlying domains. Tuning the composition to such an extreme that the metal-polymer interfacial energy becomes comparable to the metal-air interfacial energy can, of course, lead to equilibrium structures with the metal wetting one domain and dewetting from the other and forming wires. However, such an extreme is not necessary and may not even be possible with conventional polymer systems. We only need to enhance the contrast in mobilities and the energetic difference and kinetics will take care of the formation of the wires.
Another interesting aspect we presented was the presence of correlations stemming from the kinetics of the self-assembly process. We measured these correlations and showed that a simple simulation that captured the essential features of the process could indeed give correlations comparable in magnitude. Finally, we see that one could, in principle, tune the correlations by increasing or reducing the flux since cluster formation in the river regions depends on the flux.\par
Another aspect that would be useful to study is the role of grain boundaries, which has been neglected in this study. Since the wire states are polycrystalline in nature the presence of the grain boundaries will affect the stability of the structure. Another avenue of interest would be to do a more realistic simulation of the entire process, which could shed light on the distribution of sizes of nanocrystalline domains in the wires.\par
The present model appears to provide a basic understanding of the mechanism of spontaneous wire formation on striped copolymer templates. It identifies general kinetic criteria for the formation of wires. By meeting these criteria, we expect to find ways of forming such wires with a variety of metals. Further, our new understanding of the correlations has led to insight about why the wires are irregular and a strategy to reduce this irregularity.

\section{Acknowledgements}
 The author would like to thank Tom Witten, Ward Lopes, Heinrich Jaeger, Bob Gomer, Stuart Rice, Ruoyu Bao, Li-Sheng Tseng, Dan Robbins, Toan Nguyen and members of the Witten group for insightful discussions and comments. This work was supported in part by the National Science Foundation under award DMR-9975533  and via its MRSEC Program under Award Number DMR-0213745.

\end{document}